\documentclass[
twocolumn, 
showpacs, 
longbibliography, 
prx, 
aps, 
amsmath, 
amssymb,
superscriptaddress]{revtex4-2}

\usepackage{amsmath}
\usepackage{amsfonts}
\usepackage{graphicx}
\usepackage[colorlinks,linkcolor=blue,citecolor=blue,urlcolor=blue]{hyperref}
\usepackage{float}
\usepackage[normalem]{ulem}
\usepackage[dvipsnames]{xcolor}

\newcommand{\avg}[1]{\langle #1 \rangle}
\begin{document}

\title{Memory-induced Excitability in Optical Cavities}


\author{B. Braeckeveldt}
\email{Bertrand.Braeckeveldt@umons.ac.be}
\affiliation{Micro- and Nanophotonic Materials Group, Research Institute for Materials Science and Engineering, University of Mons, 20 Place du Parc, Mons B-7000, Belgium}
\author{K. J. H. Peters}
\affiliation{%
 Center for Nanophotonics, AMOLF, Science Park 104, 1098 XG Amsterdam, The Netherlands
}%
\author{B. Verdonschot}
\affiliation{%
 Center for Nanophotonics, AMOLF, Science Park 104, 1098 XG Amsterdam, The Netherlands
}%
\author{B. Maes}
\affiliation{Micro- and Nanophotonic Materials Group, Research Institute for Materials Science and Engineering, University of Mons, 20 Place du Parc, Mons B-7000, Belgium}
\author{S. R. K. Rodriguez}
\email{s.rodriguez@amolf.nl}
\affiliation{%
 Center for Nanophotonics, AMOLF, Science Park 104, 1098 XG Amsterdam, The Netherlands
}

\date{\today}

\begin{abstract}
Neurons and other excitable systems can release energy suddenly given a small stimulus. Excitability has recently drawn increasing interest in optics, as it is key to realize all-optical artificial neurons enabling speed-of-light information processing. However, the realization of all-optical excitable units and networks remains challenging. Here we demonstrate how laser-driven optical cavities with memory in their nonlinear response can sustain excitability beyond the constraints of memoryless systems. First we demonstrate different classes of excitability and spiking, and their control in a single cavity with memory. This single-cavity excitability is limited to a narrow range of memory times commensurate with the linear dissipation time. To overcome this limitation, we explore coupled cavities with memory. We demonstrate that this system can exhibit excitability for arbitrarily long memory times, even when the inter-cavity coupling rate is smaller than the dissipation rate. Our coupled-cavity system also sustains spike trains --- a hallmark of neurons --- that spontaneously break mirror symmetry.  Our predictions can be readily tested in thermo-optical cavities, where thermal dynamics effectively give memory to the nonlinear optical response. The huge separation between thermal and optical time scales in such cavities is promising for the realization of artificial neurons that can self-organize to the edge of a phase transition, like many biological systems do. 

\end{abstract}

\maketitle

\section{Introduction}


The human brain can process information with an energy efficiency that far exceeds that of digital computers. This recognition is motivating the development of brain-inspired hardware to overcome major bottlenecks in computing~\cite{lelmini18, Rios19, sebastian20, markovic_physics_2020, shastri_photonics_2021, Christensen22, Rao23}. Key to the success of this development is the realization of artificial neurons (ANs) which, like neurons in the brain, can fire information-rich energy spikes in response to small stimuli. This ability of neurons or ANs to release energy suddenly, in the form of spikes, is known as excitability.

ANs have been proposed and demonstrated on a variety of platforms~\cite{memmesheimer_designing_2006, yacomotti_fast_2006, coomans_solitary_2011, brunstein_excitability_2012, vaerenbergh_cascadable_2012, selmi_relative_2014, sengupta_hybrid_2016,  prucnal_recent_2016, shainline_superconducting_2017, khymyn_ultra-fast_2018, tait_silicon_2019, de_marinis_photonic_2019, chakravarty_supervised_2019, schneider_synaptic_2020, dunham_nanoscale_2021, hejda_resonant_2022, mei_-memory_2023}. Most state-of-the-art ANs process electrical or optical signals, and they can be made of ferroelectric~\cite{slesazeck_nanoscale_2019, prosandeev_ultrafast_2021, Noheda22, zhai_reconfigurable_2023}, phase change~\cite{chakraborty_photonic_2019, zhao_low-power_2019, joshi_accurate_2020},  two-dimensional~\cite{sun_self-selective_2019, mennel_ultrafast_2020}, or organic~\cite{van_de_burgt_non-volatile_2017, satapathi_halide-perovskite-based_2022} materials, to name a few examples. A nonlinear response is key to excitability, and hence to the realization of ANs. Many efforts have therefore concentrated on the electrical domain, where strong nonlinearities enable low-power excitability. However, recent progress in the development of highly-nonlinear photonic systems~\cite{Chang14, Togan18, Estrecho19, Mann21, Khurgin23} is making these systems increasingly attractive for the 
realization of low-power ANs. Photonic systems offer excitability at unrivalled speed, but this is not enough to mimic the brain. The brain also embraces slow dynamics to realize its remarkable information processing capability. More precisely, the coupling of slow and fast dynamical variables grants the brain an ability to self-organize to the edge of a phase transition~\cite{haldeman_critical_2005, friedman_universal_2012, diSanto16, MunozRMP, buendia_self-organized_2020, plenz_self-organized_2021, obyrne_how_2022}, where information processing is thought to be optimal and robust to parameter changes~\cite{Langton92, bertschinger_real-time_2004, boedecker_information_2012, MunozRMP, cramer_control_2020, hochstetter_avalanches_2021}.  Hence, embracing the coupling of slow and fast variables is also important for the realization of ANs and brain-inspired computation. 

In this manuscript we demonstrate how laser-driven optical cavities with memory in their nonlinear response can enable the realization of all-optical ANs. Like real neurons, our ANs embrace the  coupling of slow and fast variables to achieve excitability across a wide range of time scales, and can emit a sequence of intensity spikes known as  `spike trains'. In the brain, such spike trains encode information~\cite{borst_information_1999, diesmann_stable_1999, kumar_spiking_2010, sihn_spike_2019, de_marinis_photonic_2019, lee_photonic_2022} in both the amplitude and separation of the spikes. While our work is purely theoretical and numerical, the system and parameter range we study can be readily realized using state-of-the-art thermo-optical cavities~\cite{marino_thermo-optical_2004, yacomotti_fast_2006, vaerenbergh_cascadable_2012, fonseca_slow_2020, geng2020, peters2021}.


This manuscript is organized as follows. In Section~\ref{sec:Single} we introduce the model for a single laser-driven cavity with  memory in its nonlinear response. We demonstrate different classes of excitability and spiking in such a cavity, their connection to limit cycles, and the limited parameter range over which excitability can be achieved. Next, in Section~\ref{sec:coupled} we demonstrate how to overcome the limitations of single cavities using coupled cavities. We explore how the  coupling strength  between the cavities, the memory time, and the laser noise, impact the excitability of the coupled cavities. We show the existence of excitability for arbitrarily long memory times, and the emergence of spike trains that spontaneously break the mirror symmetry of the couple cavity system. Finally, in Section~\ref{sec:Conclusions} we provide conclusions and perspectives of our work.

\section{\label{sec:Single} Single cavity excitability}

\subsection{\label{subsec:single-nl}The model}

\begin{figure}[tbp!]
    \centering
    \includegraphics[width=\linewidth]{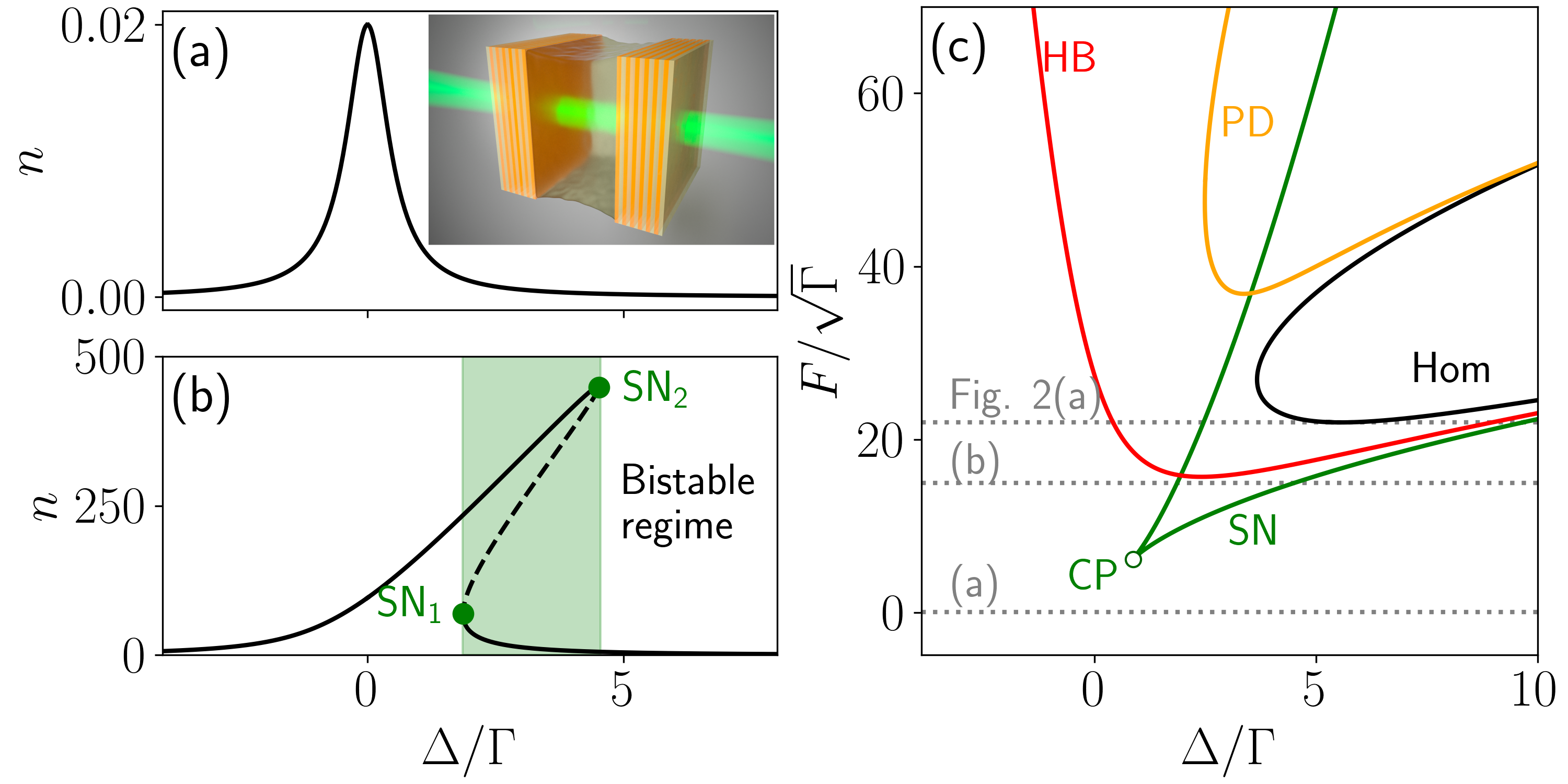}
    \caption{(a) Intensity $n=|\alpha|^2$ versus detuning $\Delta$ referenced to the dissipation rate $\Gamma$ for a laser amplitude $F=0.1\sqrt{\Gamma}$. Inset: Illustration of an oil-filled cavity described by Eq.~\eqref{eq:single_kernel}. (b) $n$ versus $\Delta/\Gamma$ for $F=15\sqrt{\Gamma}$. Solid (dashed) lines correspond to stable (unstable) solutions. The two saddle-node bifurcations (SN\textsubscript{i}) are indicated by solid green dots. Green shaded area indicates the region of bistability. (c) Bifurcations in the ($\Delta, F$)-plane. Solid curves correspond to Hopf bifurcations (HB, red), period doubling bifurcations (PD, orange), homoclinic bifurcations (Hom, black) and saddle-node bifurcation (SN, green). The open circle indicates a cusp (CP). Dashed gray lines indicate the laser amplitudes studied in Fig.~\ref{fig:single-bif}[(a),(b)] and Fig.~\ref{fig:single-bif2}(a). Parameters: $\Gamma=1$, $\kappa_\mathrm{L}=\Gamma/2$, $U=\Gamma/100$, $\tau=\Gamma^{-1}$, $D=0$.}
    \label{fig:single-bif}
\end{figure}

Consider a coherently driven single-mode resonator with memory in its nonlinear response. For concreteness, we will refer to a laser-driven plano-concave Fabry-Perot cavity filled with oil, as illustrated in Fig.~\ref{fig:single-bif}(a) and experimentally studied in Refs.~\onlinecite{geng2020, peters2021}. However, the system we envisage can also be realized using ring resonators~\cite{Lipson04, Priem05, Shi14}, whispering gallery mode resonators~\cite{Carmon04}, photonic crystal cavities~\cite{Notomi05, Brunstein09, Sodagar15, Perrier:20}, and plasmonic particles~\cite{khurgin15}, for example. All these resonators sustain a thermo-optical nonlinearity that makes the optical response non-instantaneous and leads to memory effects.

The cavity has resonance frequency $\omega_0$, intrinsic loss rate $\gamma$, and nonlinearity of strength $U$. The input-output rates through the `left' and `right' mirror are $\kappa_\mathrm{L}$ and $\kappa_\mathrm{R}$. A laser with frequency $\omega$ and amplitude $F$ drives the cavity from the left. In a frame rotating at $\omega$, the intra-cavity light field $\alpha$ follows
\begin{widetext}
\begin{equation}\label{eq:single_kernel}
    i \dot{\alpha}(t) = \left[-\Delta- i\frac{\Gamma}{2} + U \int_0^t ds\,K(t-s)|\alpha(s)|^2\right]\alpha(t) + i \sqrt{\kappa_\mathrm{L}}F + \frac{D}{\sqrt{2}} \left[\xi(t) + i \zeta(t) \right].
\end{equation}
\end{widetext}
$\Delta=\omega-\omega_0$ is the laser-cavity detuning, and ${\Gamma=\gamma+\kappa_\mathrm{L}+\kappa_\mathrm{R}}$ is the total loss rate. The term with the integral represents the non-instantaneous nonlinearity. For a thermo-optical nonlinearity as in our system, the memory kernel is  ${K(t)=\mathrm{exp}\left(-t/\tau\right)/\tau}$. The thermal relaxation time or memory time $\tau$ roughly determines how long the past exerts an influence on the system. The stochastic term ${D\left[\xi(t)+i\zeta(t)\right]/\sqrt{2}}$ accounts for Gaussian white noise with variance $D^2$ in the laser amplitude and phase. The noise components $\xi(t), \zeta(t)$ each have zero mean [i.e., ${\langle \xi(t) \rangle = \langle \zeta(t) \rangle =0}$], and are delta-correlated in time with unit variance [i.e., ${\langle \xi(t) \xi(t+t') \rangle = \langle \zeta(t) \zeta(t+t') \rangle = \delta(t')}$]. Moreover,  $\xi(t)$ and $\zeta(t)$ are mutually uncorrelated [i.e., ${\langle \xi(t) \zeta(t+t')\rangle = 0}$].

Since the kernel $K(t)$ is integrable, we can rewrite the integro-differential Eq.~\eqref{eq:single_kernel} as a set of ordinary differential equations which are convenient for numerical integration and continuation. Writing $\alpha=u + i v$ and $w=U\int_0^t ds\,K(t-s)|\alpha(s)|^2$, we obtain
\begin{equation}\label{eq:single_3D}
    \begin{split}
        \dot{u}(t) &= -\frac{\Gamma}{2} u(t) + \left[w(t)-\Delta\right] v(t) + \sqrt{\kappa_\mathrm{L}}F + \dfrac{D}{\sqrt{2}} \xi,\\
        \dot{v}(t) &= -\frac{\Gamma}{2} v(t) - \left[w(t)-\Delta\right] u(t) + \dfrac{D}{\sqrt{2}} \zeta,\\
        \dot{w}(t) &= \left\lbrace U\left[u^2(t)+v^2(t)\right] - w(t) \right\rbrace/\tau.
    \end{split}
\end{equation}

\noindent In the following, we present solutions to this set of equations. For analysing fixed points and their bifurcations under parameter continuations, we solve the deterministic equations using the software package \textsc{Auto-07p}~\cite{doedel2007}.  For analysing dynamics, we numerically integrate the stochastic equations using a homemade solver in $\textsc{Python}$.

\subsection{Fixed points and bifurcations}

First we consider the fixed points of Eq.~\eqref{eq:single_3D} for various driving conditions and without noise ($D=0$).  Equation~\eqref{eq:single_3D} has the form $\dot{\bm{\alpha}}=\bm{f}(\bm{\alpha})$, with $\bm{\alpha}=(u,v,w)^\intercal$. The fixed points satisfy $\bm{f}(\bm{\alpha})=0$, and their stability is  determined by the eigenvalues of the Jacobian of $\bm{f}$. We distinguish three types of fixed points. 1) Node: All eigenvalues are real and of the same sign. A node is stable (unstable) when all eigenvalues are negative (positive). 2) Focus: There is at least one pair of complex conjugate eigenvalues. A focus is stable when the real part of all eigenvalues is negative, and it is unstable otherwise. 3) Saddle: There is at least one real positive and one real negative eigenvalue. A saddle is always unstable.


Figure~\ref{fig:single-bif}(a) shows the intensity ${n = |\alpha|^2 = u^2+v^2}$, or intracavity photon number, versus detuning for ${F=0.1\sqrt{\Gamma}}$. For this small driving amplitude, $n$ is small, the nonlinearity is negligible, and the spectral lineshape resembles the Lorentzian of a linear cavity. In contrast, for a large driving amplitude $F=15\sqrt{\Gamma}$ the lineshape is tilted as Fig.~\ref{fig:single-bif}(b) shows. This results in a detuning range with three fixed points, of which two are stable: optical bistability. The bistable region is enclosed by two saddle-node bifurcations (SN), at which a stable point collides with a saddle and disappears~\cite{strogatz2018}. The SN bifurcations are also shown in Fig.~\ref{fig:single-bif}(c), now as a function of $F$ and $\Delta$. They are the two green curves that meet tangentially at a cusp (CP). The region enclosed by the SN bifurcations is the bistability region. 

The locations of the cusp and the saddle-nodes, as well as the optical bistability, are the same for our thermo-optical cavity and a Kerr nonlinear cavity where the nonlinearity is instantaneous. Indeed, at equilibrium, $\dot{w} = 0$ leading to $w = U n$ as for an instantaneous nonlinearity. However, the non-instantaneous nonlinearity leads to other types of dynamical state bifurcations absent in the instantaneous case. These bifurcations are due to the coupling of light to another degree of freedom, namely $w$, which gives memory to the optical response and can result in instabilities. In particular, Fig.~\ref{fig:single-bif}(c) indicates Hopf (red curve),  period-doubling (yellow curve) and homoclinic  (black curve) bifurcations, from which more complex dynamics emerge.

\subsection{\label{subsec:single-excitability}Limit cycles and excitability}

\begin{figure}[tbp!]
    \centering
    \includegraphics[width=\linewidth]{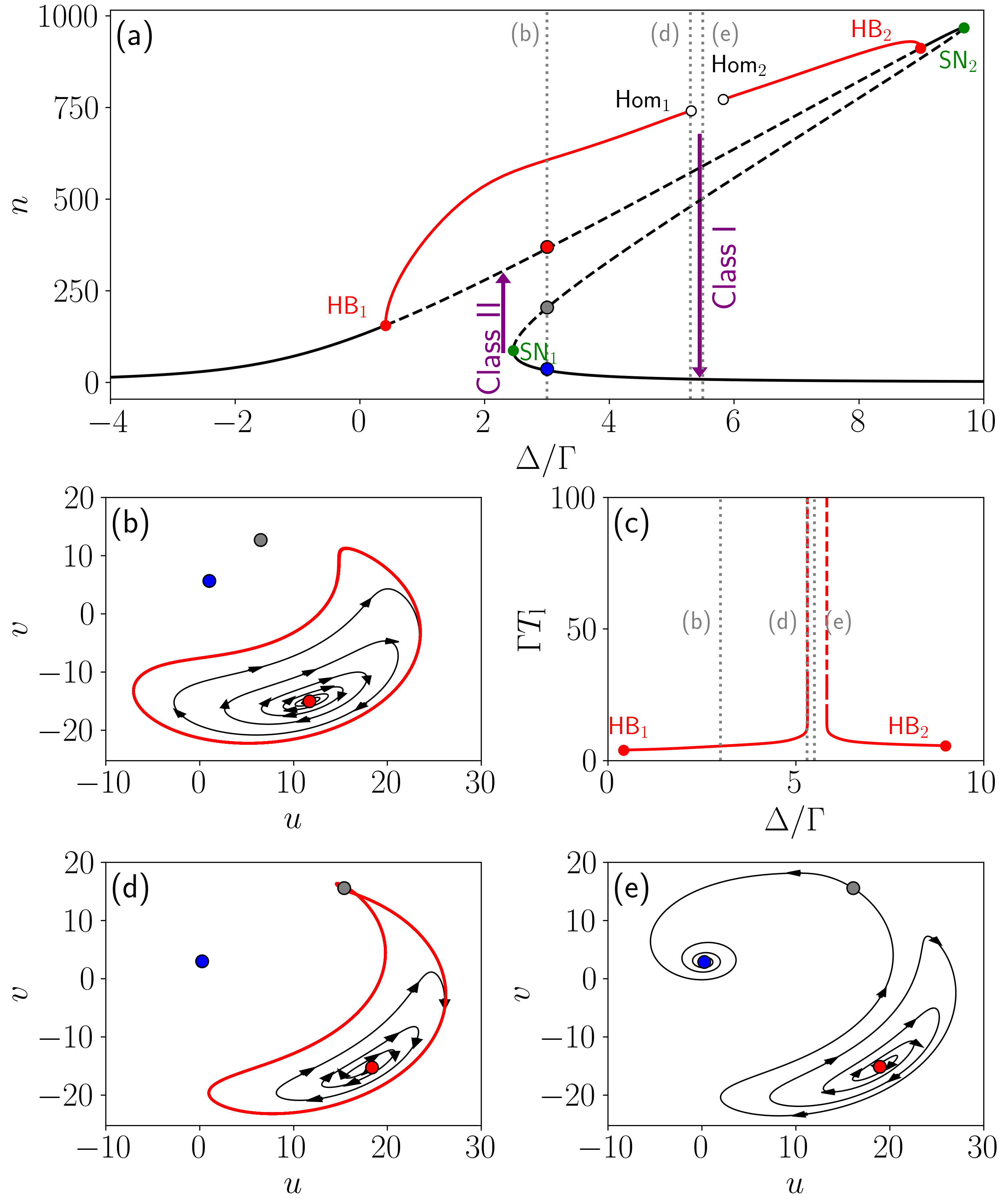}
    \caption{(a) Intensity $n$ as function of $\Delta/\Gamma$ for $F=22\sqrt{\Gamma}$. Solid (dashed) black curves correspond to stable (unstable) fixed points. Green and red solid circles indicate saddle-node (SN\textsubscript{i}) and Hopf (HB\textsubscript{i}) bifurcations, respectively. Red solid lines show the maximum $n$ of limit cycle oscillations. Open black circle indicates homoclinic bifurcations (Hom\textsubscript{i}). Purple arrows indicate transition directions with associated classes corresponding to Fig.~\ref{fig:single-class}. (b) Phase portrait corresponding to the leftmost vertical dashed gray line in panels (a) and (c). As for panels [(d)-(e)], dots indicate fixed points, with blue, red and gray corresponding to stable focus, unstable focus, and saddle, respectively. Black curves show orbits with arrows indicating the direction. Red curves show stable limit cycles. (c) Period of the limit cycle oscillations, $T_{\mathrm{l}}$ as function of $\Delta/\Gamma$. [(d)-(e)] Phase portrait corresponding to the central and rightmost dashed gray lines in panels (a) and (c). The color code is the same as for panel (b). Parameters are as in Fig.~\ref{fig:single-bif}, with $\Delta=3\Gamma$, $\Delta=5.3\Gamma$ and $\Delta=5.5\Gamma$ in (b), (d) and (e), respectively.}
    \label{fig:single-bif2}
\end{figure}

Figure~\ref{fig:single-bif2}(a) shows the intensity as a function of detuning for a large driving amplitude, $F=22\sqrt{\Gamma}$, indicated in Fig.~\ref{fig:single-bif}(c). The points labeled HB\textsubscript{1,2} are Hopf bifurcations, where a focus changes stability via a pair of purely imaginary eigenvalues~\cite{strogatz2018}. Crossing a Hopf bifurcation leads to a limit cycle --- an isolated closed orbit in phase space.  If the limit cycle is stable, self-sustained oscillations emerge. The maximum value of $n$ of these self-sustained oscillations is shown as a solid red curve in Fig.~\ref{fig:single-bif2}(a). An example of a stable limit cycle is shown in Fig.~\ref{fig:single-bif2}(b), which illustrates the phase portrait for $\Delta=3\Gamma$. Three fixed points are observed: one stable focus (blue dot), a saddle (gray dot), and an unstable focus (red dot). Trajectories starting near the unstable focus spiral outwards and converge to the limit cycle displayed in red. The system then remains on the limit cycle and self-oscillates indefinitely.

As the detuning increases from HB\textsubscript{1} to the open circle labeled Hom\textsubscript{1} in Fig.~\ref{fig:single-bif2}(a), the limit cycle grows in amplitude. The point (Hom\textsubscript{1}) is known as a homoclinic bifurcation~\cite{strogatz2018}. Approaching Hom\textsubscript{1}, the period of the limit cycle diverges [see Fig.~\ref{fig:single-bif2}(c)] as the system spends more and more time near the saddle. Exactly at the homoclinic bifurcation, a homoclinic orbit is formed: an orbit from the saddle to itself in an infinite time. This situation is depicted in Fig.~\ref{fig:single-bif2}(d), where we observe that the limit cycle has collided with the saddle (gray dot) to form a homoclinic orbit (red curve). This collision destroys the limit cycle. Thus, there are no self-sustained oscillations between Hom\textsubscript{1} and Hom\textsubscript{2}, even if the unstable focus remains. Figure~\ref{fig:single-bif2}(e) shows that no limit cycle exists for $\Delta/\Gamma=5.5$. Trajectories starting near the unstable focus spiral outwards until reaching the saddle, and then spiral towards the stable focus (blue dot). For greater detunings, the point Hom\textsubscript{2} is reached. This leads to a new limit cycle from another homoclinic bifurcation.

The above results illustrate that, unlike for a single-mode cavity with instantaneous nonlinearity~\cite{Drummond80, soljacic_optimal_2002,  khandekar_thermal_2015, Casteels16, Rodriguez17, Casteels17, peters_exceptional_2022, braeckeveldt_temperature-induced_2022, peters_scalar_2023}, a single-mode cavity with non-instantaneous nonlinearity can exhibit self-sustained oscillations. Excitability involves the transition between a resting and a spiking state, associated with stationary and fast dynamics, respectively. The system displays bistability regions with one stable focus (stationary state) and one unstable focus associated with a stable limit cycle (spiking state). At the detuning associated with the bifurcation SN\textsubscript{1} [see Fig.~\ref{fig:single-bif2}(a)] of the stable focus, there is a stable limit cycle (red solid curve). Varying a system parameter, such as the detuning in this case, can make the system transition from the resting to the spiking state. Thus we expect the system to present excitability --- the hallmark of neurons.

Different classes of excitability and spiking can be distinguished. The classification is based on the evolution of the limit cycle period $T_{\mathrm{l}}$ (spiking state) versus the free parameter~\cite{izhikevich2000} [detuning in Fig.~\ref{fig:single-bif2}(a)]. Class I: the transition between the resting state and the spiking state is accompanied by a continuous evolution of the oscillation frequency ($1/T_{\mathrm{l}}$). The oscillation period diverges at the transition. Class II: the frequency is discontinuous at the transition, and is relatively insensitive to changes in the free parameter. The frequency evolution from the resting state to the spiking state defines the class of excitability whereas the reverse transition (from spiking to resting) corresponds to the class of spiking. Next, we illustrate how both class II excitability and class I spiking can be realized in our thermo-optical cavity making it suitable for an artificial neuron.

\begin{figure}[tbp!]
    \centering
    \includegraphics[width=\linewidth]{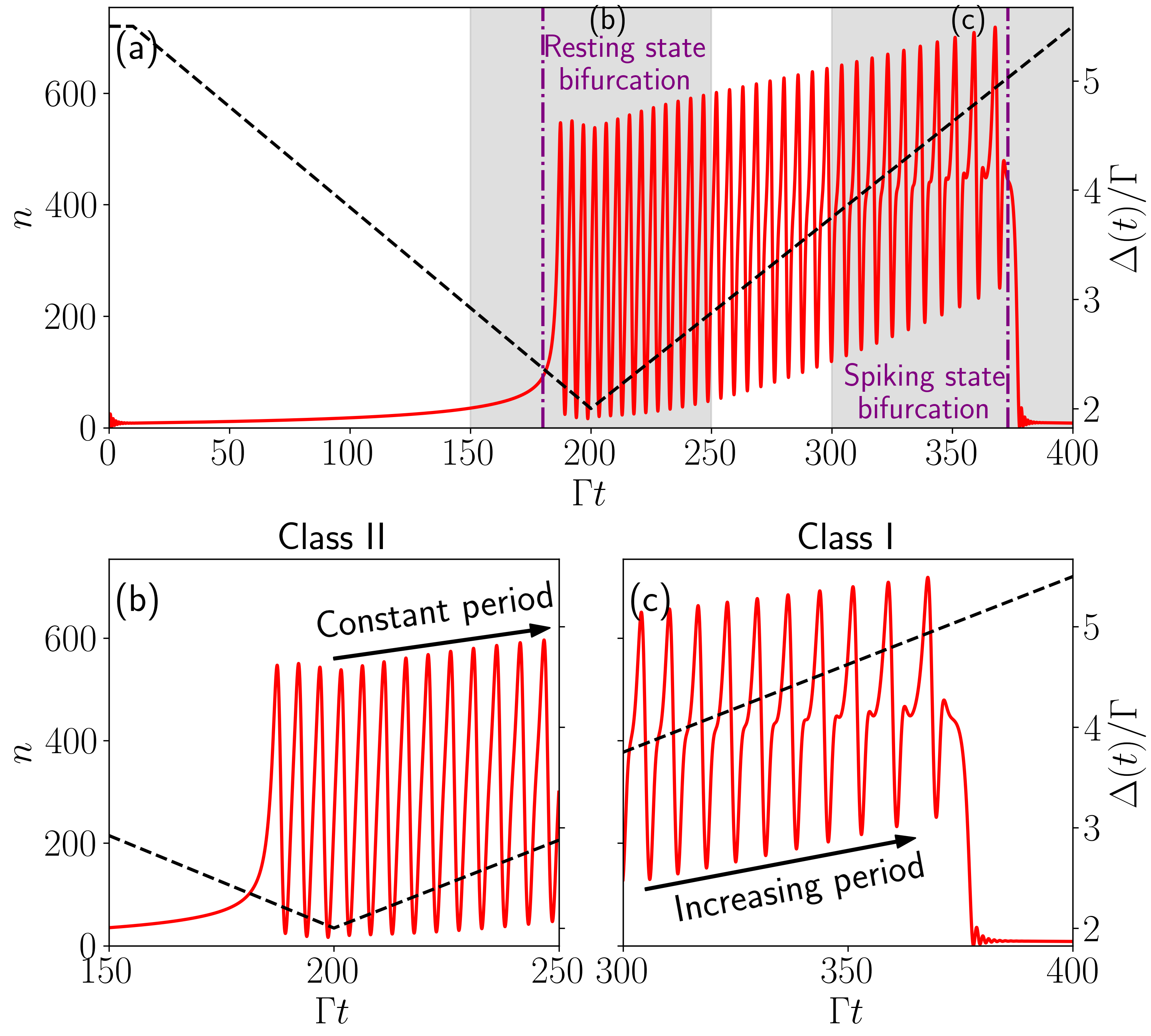}
    \caption{$n$ (solid red curve, left y-axis) and detuning $\Delta/\Gamma$ (dashed black line, right y-axis) versus time $\Gamma t$. (a) Shaded areas correspond to the close-up views [(b)-(c)] of the transition from resting to spiking states, or vice versa, located by dashed-dotted vertical purple lines. (b) Zoom into the resting state bifurcation corresponding to class II excitability in Fig.~\ref{fig:single-bif2}(a). (c) Zoom into the spiking state bifurcation associated with class I spiking in Fig.~\ref{fig:single-bif2}(a). Parameter values are as in Fig.~\ref{fig:single-bif2}(a). The detuning varies between $\Delta=5.5\Gamma$ and $\Delta=2.2\Gamma$.}
    \label{fig:single-class}
\end{figure}

Figure~\ref{fig:single-class}(a) shows $n(t)$ when $\Delta$ is ramped at fixed $F=22\sqrt{\Gamma}$, as in Fig.~\ref{fig:single-bif2}. For $\Delta = 5.5\Gamma$, the system is in a resting state (stable focus) corresponding to the low $n$ branch of Fig.~\ref{fig:single-bif2}(a). For smaller $\Delta$, this state undergoes a saddle-node bifurcation [SN\textsubscript{1} in Fig.~\ref{fig:single-bif2}(a)] when it collides with the saddle. After this bifurcation, the system is forced to enter the spiking state corresponding to the unstable focus. Thus, the system transitions from a zero frequency resting state to a near-constant frequency spiking state as displayed in Fig.~\ref{fig:single-class}(b); this corresponds to class II excitability~\cite{izhikevich2000}. The corresponding transition is represented by the upward-pointing arrow in Fig.~\ref{fig:single-bif2}(a). 

Once the system is in the oscillatory state, the unstable focus undergoes a homoclinic bifurcation when the detuning is increased. Homoclinic bifurcations are associated with divergence of the oscillation period [see Fig.~\ref{fig:single-bif2}(c)]. Thus, during the transition from the spiking to the resting state in Fig.~\ref{fig:single-class}(c), the oscillation period increases just before the transition. The corresponding transition is represented by the downward-pointing arrow in Fig.~\ref{fig:single-bif2}(a), and it is associated with class I spiking~\cite{izhikevich2000}. Therefore, the spiking state can be turned on and off via the saddle-node and the homoclinic bifurcation, respectively. 

\subsection{Limitations due to the memory time}

\begin{figure}[tbp!]
    \centering
    \includegraphics[width=\linewidth]{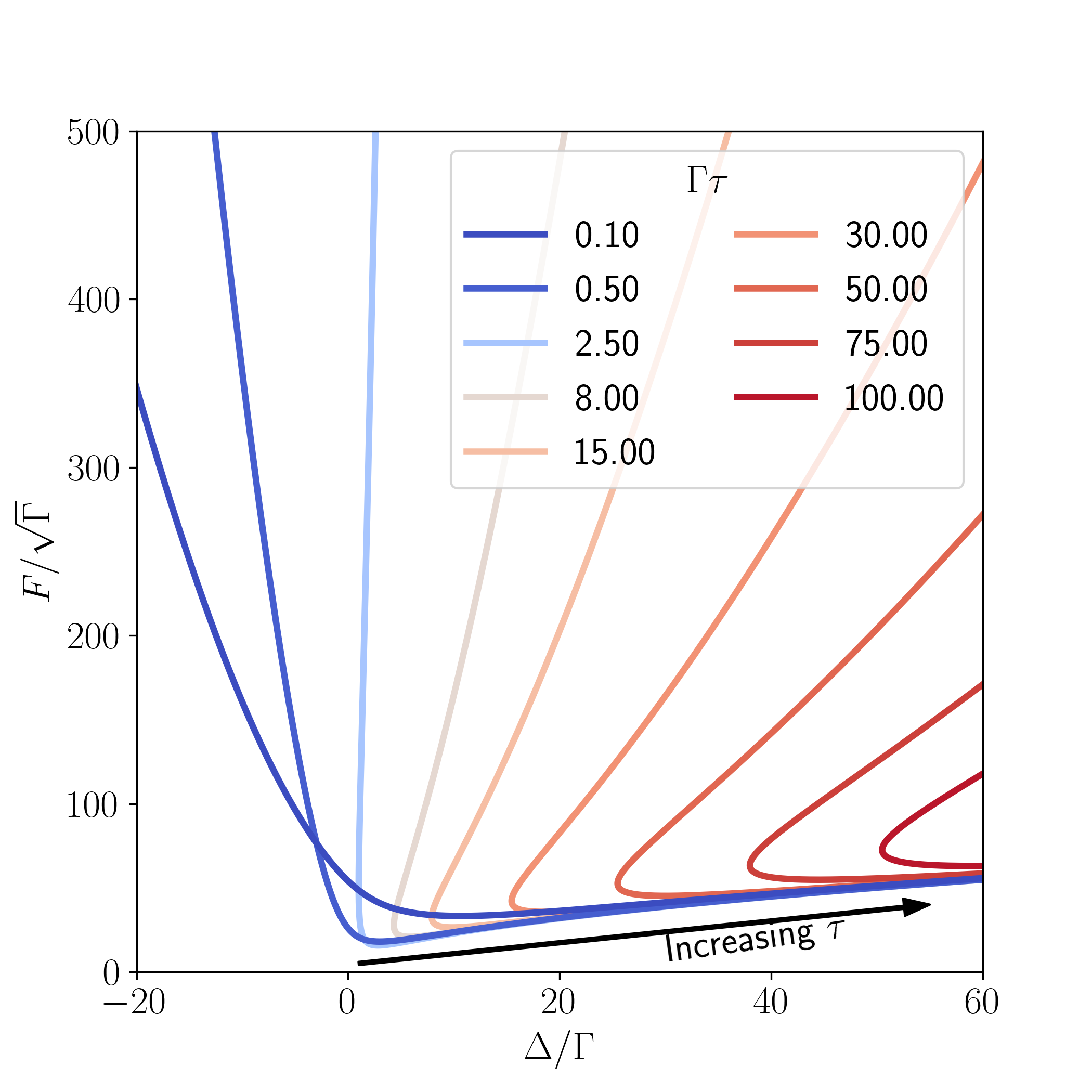}
    \caption{Hopf bifurcation curves for a single cavity in the $(\Delta,F)$-plane for various memory times. Parameters are as in Fig.~\ref{fig:single-bif}(c).}
    \label{fig:single-memory}
\end{figure}

We have so far considered a single value of the memory time: $\tau=\Gamma^{-1}$. In Fig.~\ref{fig:single-memory} we illustrate the effect of increasing $\tau$ on the Hopf bifurcations. The bifurcations shift to larger $\Delta/\Gamma$ as $\tau$ increases. This poses a major obstacle for observing, and harnessing, excitable dynamics in a single-mode cavity. The single-mode description we have made is only valid when the mode under study is well isolated, spectrally and spatially, from all other modes in the cavity. This condition can only be satisfied within a limited frequency range, which is typically less than a few tens of resonance linewidths. Consequently,  memory times larger than $\Gamma \tau \simeq 30$ would require an unfeasible separation of cavity modes to remain in the single-mode limit. Therefore, the possibility to realize limit cycles and excitable dynamics in a single-mode cavity with non-instantaneous nonlinearity seems to be limited to relatively small values of $\tau$.  Consequently, systems with typical thermo-optical nonlinearity with $\tau\sim 10^{6}\Gamma^{-1}$~\cite{geng2020,peters2021,peters2022} would be unsuitable for verifying our predictions. A solution to this problem is presented in the next section.

While memory-induced excitability in single-mode cavities may be very challenging, other systems with shorter memory times could be considered. For instance, in exciton-polariton systems it is well known that an exciton reservoir can be created  even under coherent pumping ~\cite{sarkar_polarization_2010, walker_dark_2017, stepanov_dispersion_2019, amelio_galilean_2020, Estrecho19, grudinina_dark_2022}. It turns out that the exciton reservoir acts in a way that is mathematically equivalent to a thermo-optical nonlinearity. It grants memory to the optical response, albeit on much shorter time scales.  The resulting set of equations for the system of polaritons coupled to an exciton reservoir is indeed very similar to Eq.~\eqref{eq:single_3D}, with the decay rate of the reservoir acting as a memory time. Based on parameters found in the literature~\cite{amelio_galilean_2020}, we expect such a system to present a memory time $\Gamma \tau \simeq 10$. Consequently, the exciton reservoir should lead to Hopf and homoclinic bifurcations on very short time scales. To the best of our knowledge, the effects of these bifurcations have not been observed in polariton systems. We expect this prediction to stimulate efforts in that direction, as they could reveal novel dynamical regimes so far unexplored in polariton systems.   

\section{\label{sec:coupled} Excitability of coupled cavities}

\subsection{The model}

We propose to overcome the aforementioned limitations of a single cavity by using two identical cavities linearly coupled with strength $J$. Each cavity contains a nonlinear memory term, like the single cavity of the previous section. The coupled system is described by
\begin{widetext}
\begin{equation}\label{eq:coupled_6D}
\begin{split}
\dot{u}_{j} &= - \frac{\Gamma}{2} u_{j} -\left( \Delta - w_j\right) v_{j} - J v_{3-j} + \sqrt{\kappa_\mathrm{L}} F_j + \frac{D}{\sqrt{2}} \xi_{j},\\
\dot{v}_{j} &= - \frac{\Gamma}{2} v_{j} + \left( \Delta - w_j\right) u_{j} + J u_{3-j} + \frac{D}{\sqrt{2}} \zeta_{j},\\
\dot{w}_j &= \frac{U \left( u_{j}^2 + v_{j}^2\right) - w_j}{\tau},
\end{split}
\end{equation}
\end{widetext}
with $j\in{1,2}$. The two cavities are deterministically driven with equal amplitude and phase, but the stochastic terms are independent, i.e.,
\begin{equation}
\begin{split}
    \avg{\xi_{j}(t) \zeta_{k}(t+t')} &= 0 \quad j, k \in \{1, 2\},\\
    \avg{\xi_{j}(t) \xi_{k}(t+t')} &= \delta(t') \delta_{jk} \quad j, k \in \{1, 2\}.
\end{split}
\end{equation}
We focus on dispersive coupling $J\in \mathbb{R}^+$, which is typical of experimental systems~\cite{hamel_spontaneous_2015, rodriguez_interaction-induced_2016, carlon_zambon_parametric_2020, xu_spontaneous_2021, garbin_spontaneous_2022, Toebes22}. For $J>\Gamma$, the spectrum comprises a symmetric  ($\alpha_1=\alpha_2$) and antisymmetric ($\alpha_1=-\alpha_2$) superposition of the bare cavity modes, split by $2J$.

Even without memory ($\tau \rightarrow 0$), coupled cavities can sustain complex dynamics which have drawn great interest in recent years~\cite{maes_switching_2006, sarchi_coherent_2008, maes_self-pulsing_2009, rodriguez_interaction-induced_2016, hendry_spontaneous_2018, nielsen_coexistence_2019, giraldo_driven-dissipative_2020, xu_spontaneous_2021, yelo-sarrion_self-pulsing_2021, yelo-sarrion_self-pulsing_2022, yelo2022, braeckeveldt_thermal_2023, heugel_role_2023}. Hopf and homoclinic bifurcations and chaos~\cite{sarchi_coherent_2008, maes_self-pulsing_2009, giraldo_driven-dissipative_2020, yelo-sarrion_self-pulsing_2021}, as well as excitability~\cite{yelo2022}, have been numerically shown in the strong coupling regime ($J > \Gamma$). In all those systems, limit cycles have a relatively short period  ${T_\mathrm{l} \propto J^{-1} \sim}$~ps limited by the high coupling rates of optical cavities. The sub-ps time resolution needed to observe such oscillations has made their direct experimental observation and utilization impossible so far. 

In the following, we show that coupled cavities with non-instantaneous nonlinearity  can overcome the limitations of both the single-cavity system with non-instantaneous nonlinearity, as well as those of coupled-cavity systems with instantaneous nonlinearity. In particular, we will show that limit cycles and excitability can be realized in systems with memory times $\tau$ much larger than the dissipation time  $\Gamma^{-1}$. 


\subsection{Fixed points and bifurcations}

\begin{figure}[tbp!]
    \centering
    \includegraphics[width=\linewidth]{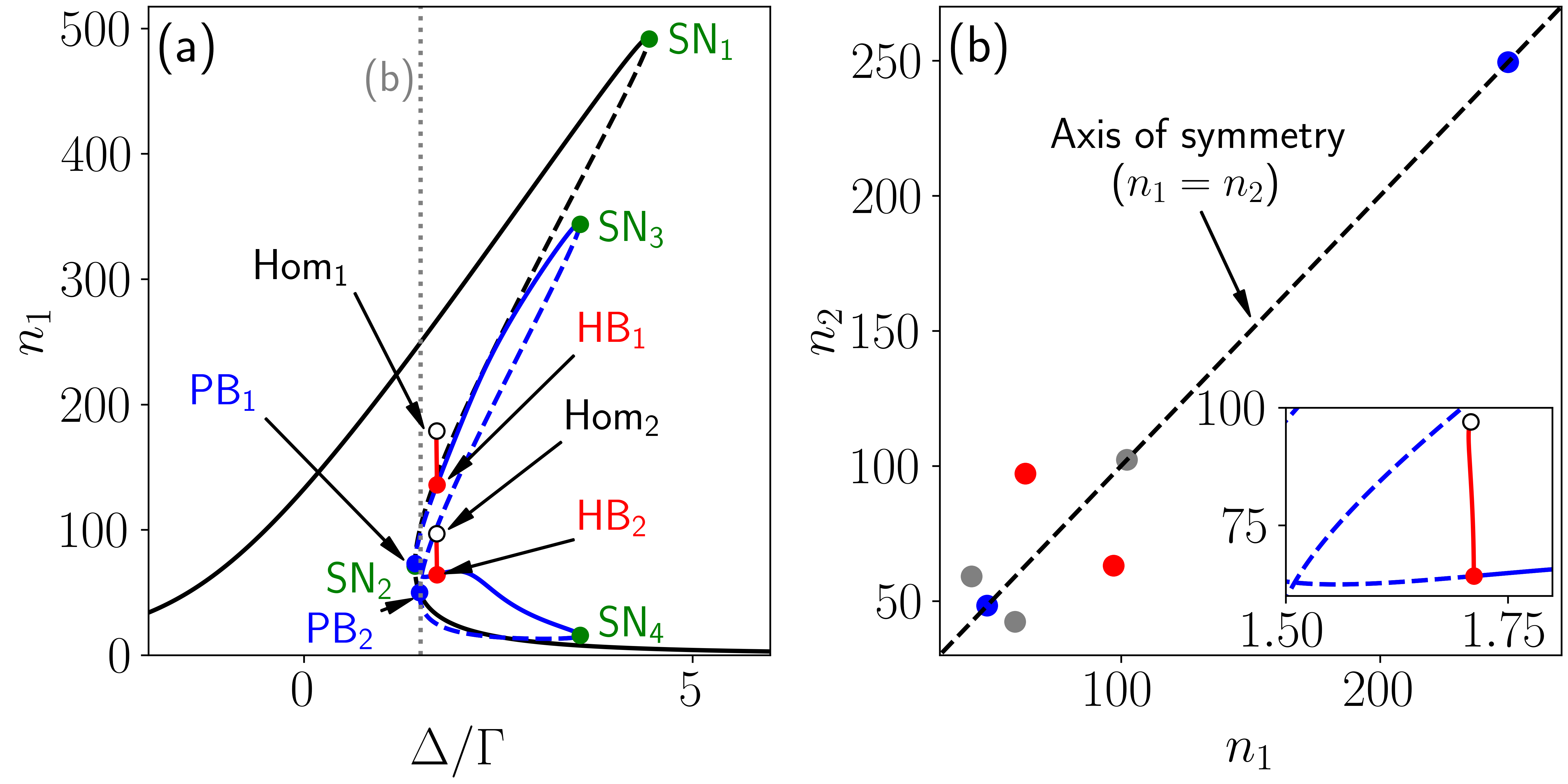}
    \caption{(a) Bifurcation diagram of $n_1$ along  the detuning $\Delta$ for coupled and symmetrically driven cavities. Solid (dashed) curves represent stable (unstable) fixed points with symmetric and asymmetric solutions in black and blue, respectively. Saddle-node (SN, green), Hopf (HB, red), homoclinic (Hom, black) and pitchfork (PB, blue) bifurcations are located with green, red, empty, and blue dots, respectively. (b) Fixed points for a detuning $\Delta=1.5\Gamma$ color coded as in Fig.~\ref{fig:single-bif2}. The dashed black line shows the axis of symmetry $n_1=n_2$. Inset: close-up view of Hopf (HB\textsubscript{1}) and homoclinic (Hom\textsubscript{1}) bifurcations. The red solid curve represents the maximum amplitude of the stable limit cycle. Parameters: $\Gamma = 1$, $U = \Gamma/100$, $\kappa_\mathrm{L} = \Gamma/2$, $D = 0$, $J=0.5\Gamma$, $F=15.7\sqrt{\Gamma}$, $\tau=100\Gamma^{-1}$.}
    \label{fig:coupled-bif1}
\end{figure}

Coupled cavities support fixed points absent in the single cavity. In Fig.~\ref{fig:coupled-bif1}(a) we plot the intensity in cavity 1 ($n_1$) versus $\Delta/\Gamma$. We obtained this plot by solving Eq.~\eqref{eq:coupled_6D} with constant driving amplitude [$F=15.7\sqrt{\Gamma}$]. Solutions are color coded as in the previous section. The black curves correspond to the symmetric solutions and behave similarly to the single-cavity [see Fig.~\ref{fig:single-bif}(b)] due to the symmetric driving. Between saddle-node bifurcations labeled SN\textsubscript{1,2}, the coupled system exhibits three symmetric states, two of them being stable and one unstable. Asymmetric solutions [blue branches in Fig.~\ref{fig:coupled-bif1}(a)] are connected to the symmetric one via pitchfork bifurcations (PB\textsubscript{1,2}), indicated by the blue dots in Fig.~\ref{fig:coupled-bif1}(a). The location of these bifurcations can be computed analytically (see Appendix~\ref{app:pitchfork}).

To illustrate the complex behavior appearing in the multi-stability region, Fig.~\ref{fig:coupled-bif1}(b) shows the fixed points in the ($n_1, n_2$) space for $\Delta=1.5\Gamma$, color coded as indicated in the caption. Due to the mirror symmetry, the intensities in both cavities can be expected to be equal and therefore lie along the symmetry axis indicated by the dashed black line. This mirror symmetry can spontaneously break~\cite{cao_two_2016, garbin_spontaneous_2022} and asymmetric solutions  $n_1\neq n_2$ emerge.  Interestingly, limit cycles appear in asymmetric solutions as they undergo Hopf bifurcations (HB\textsubscript{1,2}) represented by red dots. A stable limit cycle with a maximum intensity displayed via a short red line emerges from each of these bifurcations. Limit cycles disappear via homoclinic bifurcations (Hom\textsubscript{1,2}) located with open circles. A zoom in these bifurcations is shown as an inset in Fig.~\ref{fig:coupled-bif1}(b). 

Remarkably, all bifurcations in Fig.~\ref{fig:coupled-bif1}(a) correspond to a small coupling $J=0.5\Gamma$ and large memory time $\tau=100/\Gamma$. This already shows that the limitations previously discussed are gone. Coupled cavities with non-intantaneous nonlinearity can sustain limit cycles for very long memory times, even when the coupling is weaker than the dissipation.



Figure~\ref{fig:coupled-bif1}(a) shows that Hopf bifurcations occur in the asymmetric branches, absent in the single cavity system. To relate these results to single cavity physics, let us assume excitation of the symmetric mode by setting $\alpha_1=\alpha_2$ in Eq.~\eqref{eq:coupled_6D}. By considering steady states without noise ($D=0$), one obtains
\begin{equation}\label{eq:sym-solutions}
0 = \left[i \left( \Delta + J - U |\alpha_j|^2\right) - \dfrac{\Gamma}{2} \right] \alpha_j +  \sqrt{\kappa_\mathrm{L}} F,
\end{equation}
with $j \in {1, 2}$. The above equation is equivalent to that describing a single cavity, but  with the detuning shifted by the coupling $J$. Taking the square modulus leads to a cubic equation in $n_j=|\alpha_j|^2$, which gives three possible symmetric solutions (see Appendix~\ref{app:symmetric-solution}). However, these symmetric solutions are not the only ones. Even under symmetric driving, asymmetric solutions emerge due to spontaneous symmetry breaking. One can show (see Appendix~\ref{app:asym-conditions}) that the steady states generally satisfy
\begin{equation}\label{eq:asym-solutions}
    \begin{split}
        (n_1 - n_2) \bigg[&U^2 (n_1^2 + n_2^2 + n_1 n_2)\\
        &- 2 U (\Delta - J) (n_1 + n_2) + \dfrac{\Gamma^2}{4}\\
        &+ \left( \Delta - J\right)^2 \bigg] = 0.
    \end{split}
\end{equation}

Equation~\eqref{eq:asym-solutions} is the product of two terms, and it is therefore satisfied if at least one of them is zero. The first leads to symmetric solutions because it is zero for $n_1=n_2$. Consequently, there is always at least one symmetric solution. The second term (in the square brackets) can cancel even for asymmetric solutions $n_1 \neq n_2$. These asymmetric solutions exist under conditions (see Appendix~\ref{app:asym-conditions})
\begin{subequations}
    \begin{eqnarray}
    \Delta &\geq& J + \dfrac{\sqrt{3}}{2} \Gamma, \label{eq:detuning_condition}\\
    n_j &\leq& \dfrac{2 (\Delta - J) + \sqrt{4 (\Delta - J)^2 - 3 \Gamma^2}}{3 U}, \label{eq:n1_condition1}\\
    n_j &\geq& \dfrac{2 (\Delta - J) - \sqrt{4 (\Delta - J)^2 - 3 \Gamma^2}}{3 U}.\label{eq:n1_condition2}
    \end{eqnarray}
\end{subequations}

\noindent Due to the mirror symmetry of the system, asymmetric solutions always appear in pairs, with $n_1$ and $n_2$ symmetric around the axis $n_1=n_2$ [Fig.~\ref{fig:coupled-bif1}(b)]. There can be 2, 4, 6, or 8 asymmetric solutions. Figure~\ref{fig:coupled-bif1}(a) and the inset of Fig.~\ref{fig:coupled-bif1}(b) show that, unlike the branches associated with the symmetric solutions, the asymmetric branches support Hopf and homoclinic bifurcations. Stable limit cycles exist in between these bifurcations and are therefore present for a small detuning range. 

\subsection{Limit cycles and excitability}

The coupled cavity system also displays class I spiking and class II excitability, as the system can transition from a symmetric stable focus to an asymmetric unstable focus associated with a stable limit cycle. To induce the symmetry breaking we need an initial slight asymmetry in the driving fields. We therefore define the driving amplitude imbalance between the two cavities $\rho=(F_1 - F_2)/(F_1+F_2)$ and we vary this parameter to reveal excitable dynamics. 

\begin{figure}[tbp!]
    \centering
    \includegraphics[width=\linewidth]{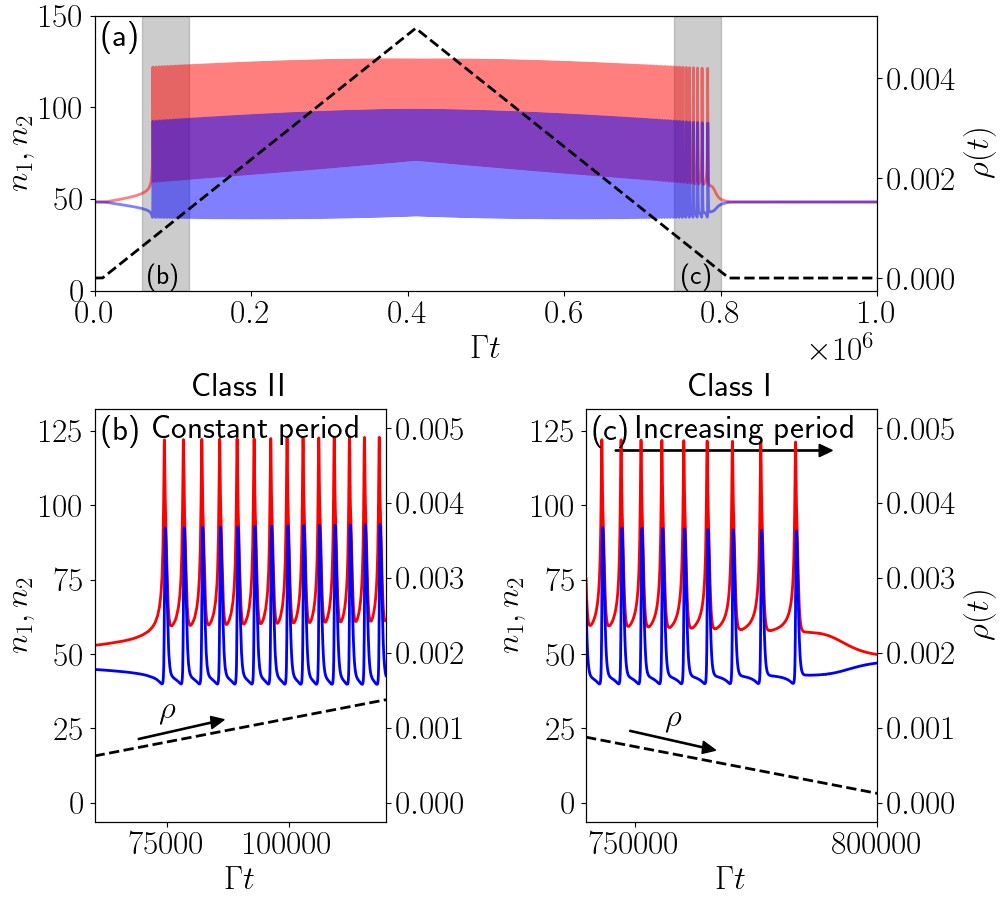}
    \caption{$n_1$ (solid red curve, left y-axis), $n_2$ (solid blue curve, left y-axis) and power imbalance $\rho=(F_1-F_2)/(F_1+F_2)$ (dashed black, right y-axis) versus time $\Gamma t$. (a) Shaded areas correspond to the close-up views [(b)-(c)] of the transition from resting to spiking states, or vice versa. (b) Zoom into the transitions from resting to the spiking state associated with class II excitability. (c) Zoom into the transition from the spiking to the resting state corresponding to class I spiking. The average power $F=[F_1(t)+F_2(t)]/2$ is constant and equals to $F=15.7\sqrt{\Gamma}$. The detuning $\Delta/\Gamma=1.5$. Other parameters are as in Fig.~\ref{fig:coupled-bif1}(a), the power imbalance $\rho$ varies between $0$ and $5\times 10^{-3}$.}
    \label{fig:coupled-class}
\end{figure}

Figure~\ref{fig:coupled-class}(a) shows the time evolution of  $n_1$ and $n_2$  when the power imbalance $\rho$ is ramped. We observe two transitions in the shaded areas of Fig.~\ref{fig:coupled-class}(a). The first transition, displayed in Fig.~\ref{fig:coupled-class}(b), shows the jump from a symmetric resting state to an asymmetric stable limit cycle. The oscillation period goes from zero to a near-constant value and is therefore discontinuous and characteristic of class II excitability. In contrast, the other transition presented in Fig.~\ref{fig:coupled-class}(c) shows that the intensities switch from an asymmetric stable limit cycle to a symmetric rest state. The transition is associated with a continuous increase of the oscillations period associated with class I spiking. Again, the coupled system presents a spiking dynamics that can be switched on or off. In comparison to the single cavity, the oscillatory state  is triggered by varying the power imbalance instead of the detuning.

\subsection{Persistence and accessibility of excitability}

\begin{figure}[tbp!]
    \centering
    \includegraphics[width=\linewidth]{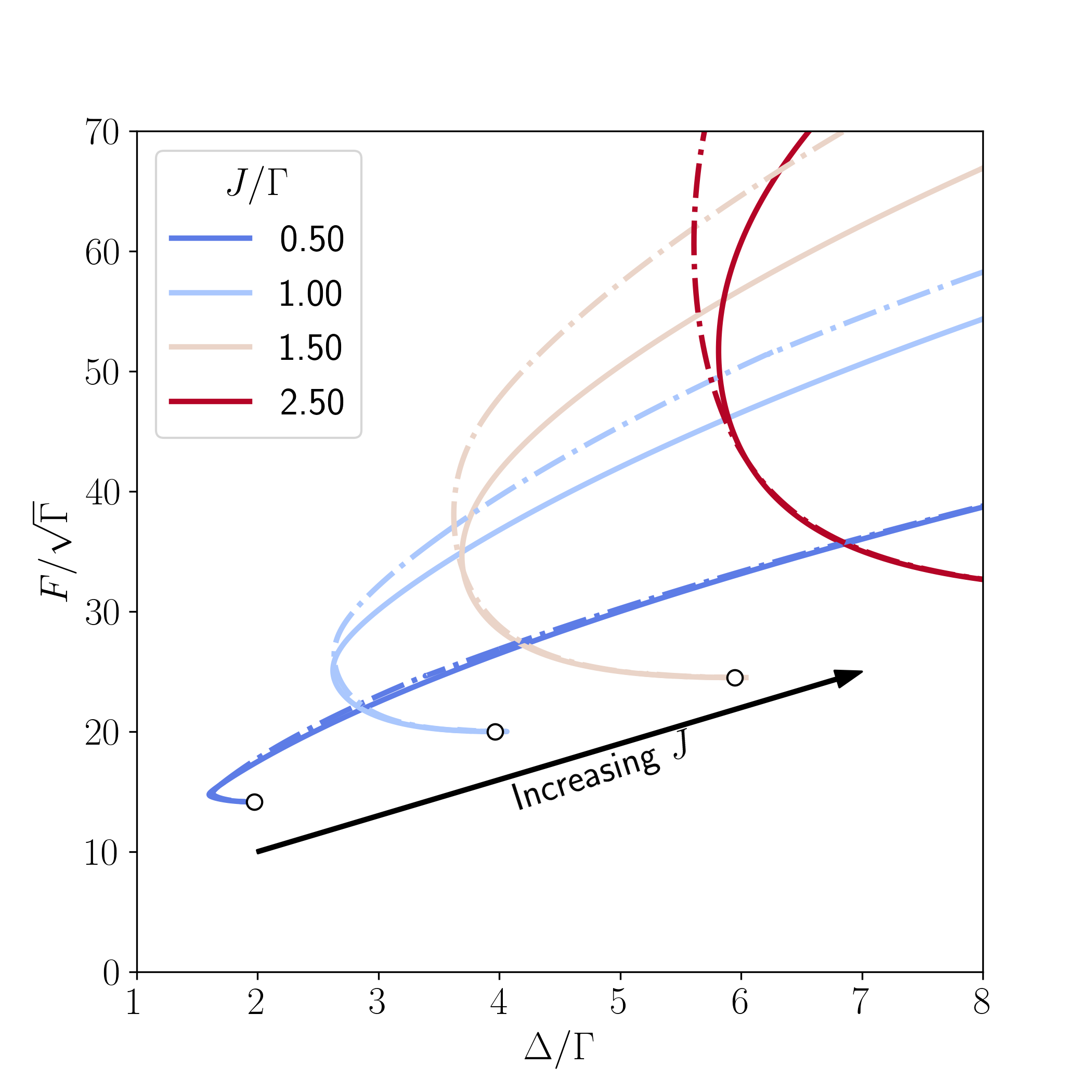}
    \caption{Hopf (solid curves) and homoclinic (dash-dotted curves) bifurcations in the $(\Delta, F)$-space for various inter-cavity coupling strengths $J$. White dots indicate Bogdanov-Takens (BT) bifurcations. Parameters are as in Fig.~\ref{fig:coupled-bif1}.}
    \label{fig:coupled-coupling}
\end{figure}

In Fig.~\ref{fig:coupled-coupling} we investigate the role of the coupling on the existence of limit cycles in our coupled cavity system.  We fix all parameters, except $J$, to the values used in Fig.~\ref{fig:coupled-bif1} The region where limit cycles exist is located between the solid and dash-dotted curves of the same color, delimiting Hopf and homoclinic bifurcations, respectively. Open circles indicate Bogdanov-Takens bifurcations, where saddle-node, homoclinic and Hopf bifurcations intersect. For a weak coupling ($J=0.5\Gamma$), oscillations occur in a narrow detuning range  as we have seen in Fig.~\ref{fig:coupled-bif1}. Interestingly,  these limit cycles persist for much smaller detuning values than in the single cavity system, even though the memory time is relatively large. Previously, in Fig.~\ref{fig:single-memory}, we showed that $\Gamma \tau = 100$ would require $\Delta > 50\Gamma$ to observe limit cycles in a single cavity. For such extremely large detunings, no experiment is likely to be well described by a single-mode model. In contrast, for coupled cavities and $\Gamma \tau = 100$, limit cycles appear around $\Delta \simeq 1.6\Gamma$ which is experimentally feasible. 

As $J$ increases, Hopf and homoclinic bifurcations  shift to higher driving amplitudes and detuning. This can be expected by noting that Hopf bifurcations appear in asymmetric branches. We noted that these solutions appear for detunings satisfying Eq.~\eqref{eq:detuning_condition}, leading to larger $\Delta$ for an increasing coupling $J$. Moreover, the intensities $n_{1,2}$ also increase according to Eqs.~[\eqref{eq:n1_condition1}-\eqref{eq:n1_condition2}], which means large driving amplitudes $F/\sqrt{\Gamma}$ are needed. Nonetheless, even for the larger coupling ${J=2.5\Gamma}$, the driving amplitude and detuning for limit cycles can still be achieved experimentally. For energy-efficient ANs, weak couplings $J$ are better as they enable spiking activity with lower laser powers.

\subsection{Excitability for arbitrarily long memory times}

\begin{figure}[tbp!]
    \centering
    \includegraphics[width=\linewidth]{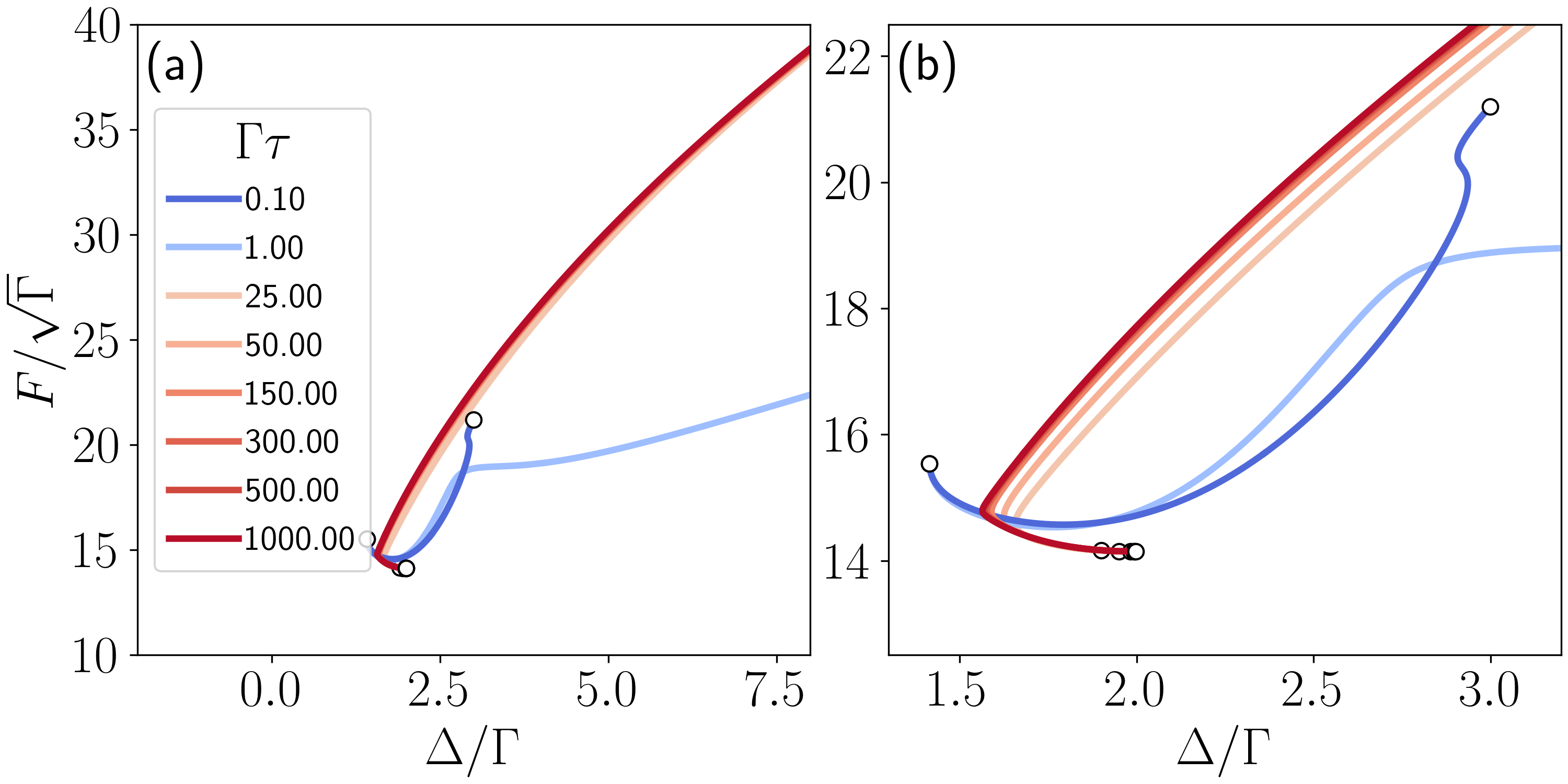}
    \caption{(a) Hopf bifurcation on the asymmetric branches in the $(\Delta, F)$-space for various memory times $\tau$ when the coupling $J=0.5\Gamma$. (b) Close-up view on the limit behavior when $\Gamma \tau \gg 1$. Parameters are as in Fig.~\ref{fig:coupled-bif1}.}
    \label{fig:coupled-memory}
\end{figure}

Figure~\ref{fig:coupled-memory} illustrates how the Hopf bifurcation in the asymmetric branch depends on the memory time $\tau$. We use the same parameters as in Fig.~\ref{fig:coupled-coupling}, except that the coupling is now fixed at $J=0.5\Gamma$ and  $\tau$ varies. To avoid an overcrowded figure, we display Hopf and not    homoclinic bifurcations. Open circles  indicate Bogdanov-Takens bifurcations. For memory times $ \Gamma \tau > 25$, the location of the Hopf bifurcation is not significantly influenced by $\tau$  as shown in Fig.~\ref{fig:coupled-memory}(a). The homoclinic bifurcation behaves similarly as the Hopf bifurcation with the increase of the memory time.  In Fig.~\ref{fig:coupled-memory}(b) we provide a close-up view of the same curves, where we see that they all merge for $\Gamma \tau > 150$. Thanks to this limiting behavior for $\Gamma \tau \gg 1$, effects of the Hopf bifurcation could be observed even in systems with extremely large memory times, such as the oil-filled cavities in Refs.~\onlinecite{geng2020, peters2021}.

When the memory and dissipation times are commensurate ($ \tau \sim \Gamma^{-1}$), the Hopf bifurcation curve has a different shape. As for a single cavity, the interplay of   dissipation  and  memory effects  leads to more complex dynamics, and many fixed points become unstable foci.

Next we demonstrate that increasing $\tau$ leads to proportionally longer limit cycles, thus enabling `slow firing'. Via a modification of the memory time, the firing period can be tailored. The behaviour we discuss is similar to spike frequency adaptation, also known as spike accommodation, observed in neurons~\cite{ha_spike_2017}. Spike frequency adaptation is associated with a decrease in the frequency of spiking during an extended period of excitation. This frequency decrease is related to a slow recovery of the calcium-activated potassium channel responsible for the action potential~\cite{benda_universal_2003}.

\begin{figure}[tbp!]
    \centering
    \includegraphics[width=\linewidth]{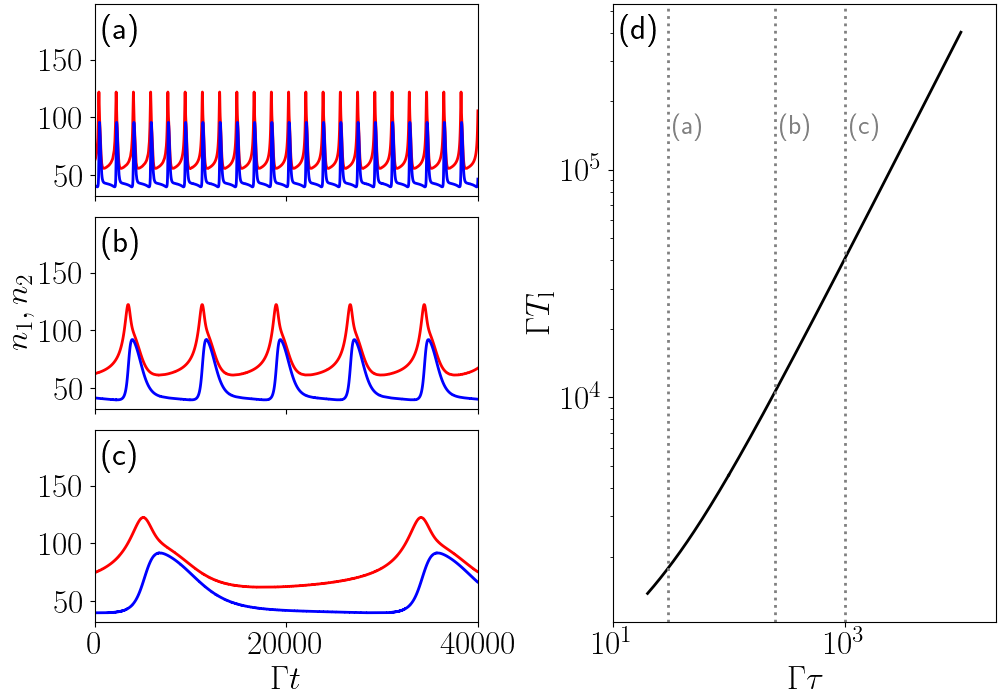}
    \caption{[(a)-(c)] $n_1$ (red) and $n_2$ (blue) versus time in a limit-cycle regime for different memory times $\tau$. (a) $\tau=30\Gamma^{-1}$. (b) $\tau=250\Gamma^{-1}$. (c) $\tau=1000\Gamma^{-1}$. (d) Continuation of the limit cycle period $T_{\mathrm{l}}$ versus memory $\tau$ (solid black curve). The dotted vertical lines show the memory time values corresponding to (a)-(c). $F_1=15.72\sqrt{\Gamma}$, $F_2=15.68\sqrt{\Gamma}$. Other parameters are as in Fig.~\ref{fig:coupled-class}.}
    \label{fig:coupled-period}
\end{figure}

Figure~\ref{fig:coupled-period} illustrates how the period of the limit cycle depends on $\tau$. Figures~\ref{fig:coupled-period}(a), ~\ref{fig:coupled-period}(b), and ~\ref{fig:coupled-period}(c) correspond to $\Gamma \tau =$ 30, 250, and 1000, respectively. Red and blue curves represent intensities in cavities 1 and 2, respectively. To set the system in the observed asymmetric self-oscillating states, we initialise our simulations close to the unstable focus associated with the stable limit cycle. Then, as the memory time increases, the period of self-pulsing increases [Figs~\ref{fig:coupled-period}(a)-(c)]. 

Figure~\ref{fig:coupled-period}(d) shows an approximately linear relation between the limit cycle period $T_{\mathrm{l}}$ and the memory time~$\tau$. We note that this log-log scale curve is a continuation calculation (using Auto-07p) along the memory time of the fixed point associated with the stable limit cycle. Importantly, this results evidences that, unlike in previous systems (single cavity with non-instantaneous nonlinearity, and coupled cavities with instantaneous nonlinearity), limit cycles with periods unconstrained by $J$ can emerge. These limit cycles can be realized for small couplings and small driving powers, as in Figs.~\ref{fig:coupled-class}~and~\ref{fig:coupled-period}, and for arbitrarily long memory times. Since both Hopf and homoclinic bifurcations shift to larger driving powers as the inter-cavity coupling increases (see Fig.~\ref{fig:coupled-coupling}), the laser power needed for excitability increases with the  coupling strength. Consequently, weakly coupled cavities with long memory times are promising for the realization of low-power ANs. Moreover, the possibility to have coupled slow and fast variables (separated by many orders of magnitude) in this system, opens the possibility of self organization to the edge of a phase transition, as thought to occur in the brain~\cite{buendia_self-organized_2020, plenz_self-organized_2021}.

\subsection{\label{subsec:coupled-noise}Noise-induced spike trains}
\begin{figure}[tbp]
    \centering
    \includegraphics[width=\linewidth]{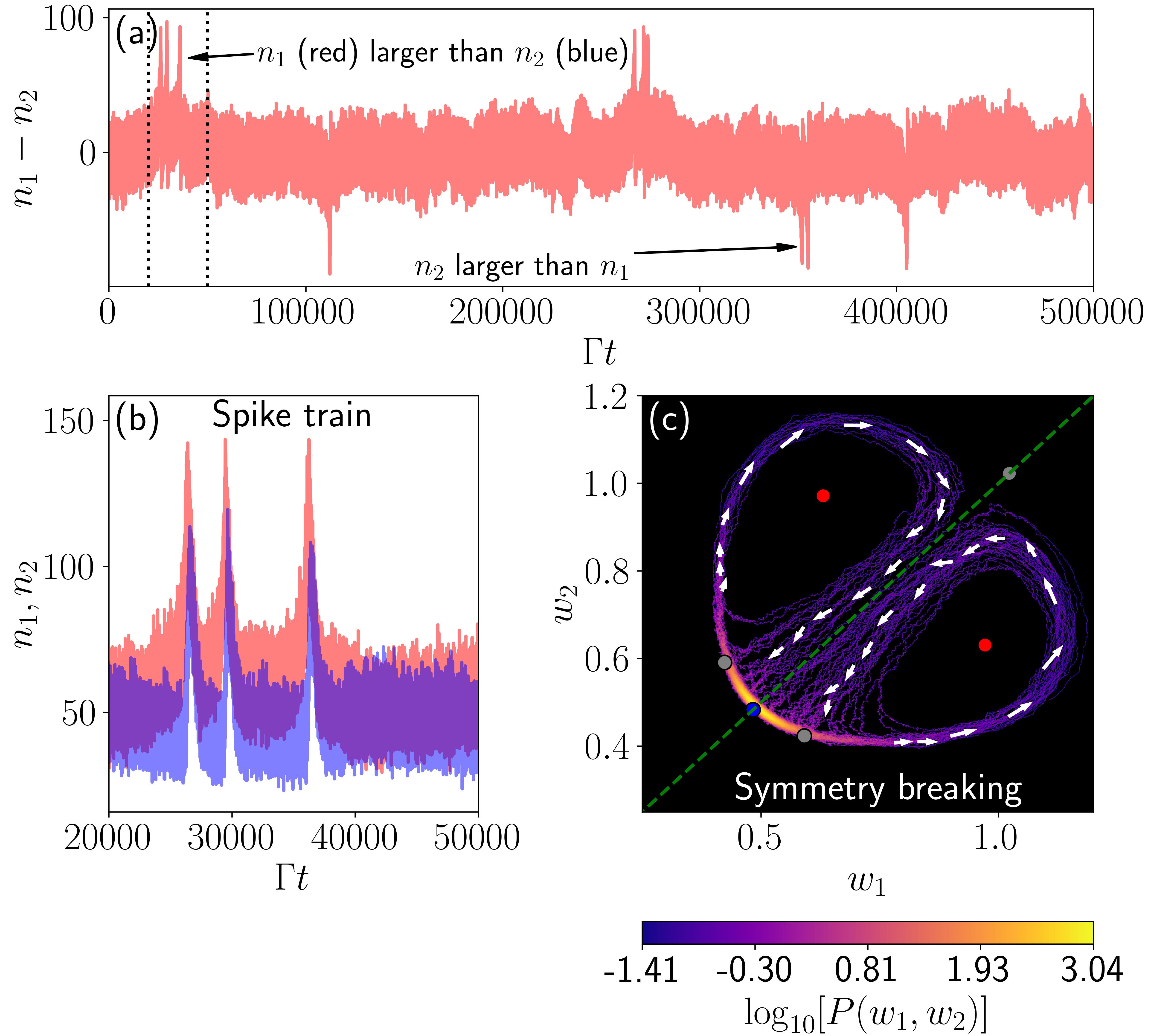}
    \caption{Stochastic dynamics of the intensities $n_1$ and  $n_2$, for a standard deviation of the laser noise $D=0.55\sqrt{\Gamma}$. (a)~Intensity difference $n_1-n_2$ versus time. (b) Zoom in the first bursting event of (a) with $n_1$ in red and $n_2$ in blue. (c) Probability density function $P(w_1, w_2)$ computed from numerical time traces. Small white arrows indicate the rotation direction of each stable limit cycle. Dots represent fixed points color coded as in Fig.~\ref{fig:single-bif2}. The dashed green line shows the axis of symmetry ($w_1 = w_2$). Parameters are as in Fig.~\ref{fig:coupled-class} with $F=15.7\sqrt{\Gamma}$ and $\rho=0$.}
    \label{fig:coupled-spiking}
\end{figure}

So far we have only considered the behavior of our system in the absence of noise. In this section we investigate how noise, in the laser amplitude and phase, results in asymmetric spike trains even when the deterministic driving is symmetric.  Figure~\ref{fig:coupled-spiking}(a) shows the intensity imbalance in the two cavities, $n_1 - n_2$, versus time. The driving is symmetric on the two cavities, with amplitude $F=15.7\sqrt{\Gamma}$. The detuning is $\Delta=1.5\Gamma$, as in Fig.~\ref{fig:coupled-class}. Even though the standard deviation of the noise is the same for both cavities ($D=0.55\sqrt{\Gamma}$), small asymmetries can be generated because the noise sources are independent.  In the calculation, we set  $\Gamma \tau = 100$ to avoid unnecessarily long computation times associated with long memory times.  However, in view of the already large separation of time scales between  $\Gamma^{-1}$ and $\tau$, qualitatively similar dynamics is expected for slower systems. 

In Fig.~\ref{fig:coupled-spiking}(a)  the system starts in a symmetric rest state, corresponding to the state at $t{> 8\times10^5 \Gamma^{-1}}$ in Fig.~\ref{fig:coupled-class}. Then, it suddenly and briefly transitions to symmetry broken states associated with spiking as in Fig.~\ref{fig:coupled-class}(b). Figure~\ref{fig:coupled-spiking}(b) shows a close-up view of the interval containing the first three spikes in this particular trajectory, with $n_1$ and $n_2$ now plotted separately. Initially, the two intensities fluctuate around the same value indicating symmetric behavior. Then the symmetry is broken during the spikes due to the random imbalance in the driving. 

The symmetry breaking is easier to observe in the $(w_1, w_2)$-space by looking at the probability density function (PDF) $P(w_1, w_2)$ as in Fig.~\ref{fig:coupled-spiking}(c). The PDF is computed from 8 trajectories of duration $5\times 10^5/\Gamma$, each associated with $2 \times 10^6$ time steps. The probability is largest close to the symmetric stable focus indicated by the blue dot. The probability decreases approaching the two saddles (gray dots) close to the stable focus. When the system reaches a saddle, it undergoes a saddle-node bifurcation and jumps to the nearest unstable focus (red dot), initiating orbits around it. The symmetry broken states are associated with limit cycles. The random fluctuations can then force the system to return to the stable focus. The transition from the symmetry-broken spiking states to the symmetric rest state can also be induced by a homoclinic bifurcation when the limit cycle swells and forms a homoclinic orbit.

In Fig.~\ref{fig:coupled-spiking}(c)  we indicate the directions of the orbits around the two unstable foci with white arrows. There are two possible orbits because there are two symmetry-broken states (red dots) associated with spiking activity. Each of these states corresponds to a different sign of the power imbalance $\rho$. A positive $\rho$ leads to $n_1$ larger than $n_2$ while the situation is reversed when $\rho$ is negative. Here, the noise terms acting on each cavity are uncorrelated leading to small imbalances on the total driving amplitudes. In other words, the combined noise terms  effectively act as a fluctuating $\rho$. Thus, the imbalance $\rho$ fluctuates between positive and negative values with equal probability, leading to spontaneous switching between both asymmetric spiking states. This enables the system to switch from one limit cycle to another, spontaneously. While spontaneous symmetry breaking effects have been observed for localized modes~\cite{Malomed13, Hamel15, Cao17, Garbin20, Xu21, Garbin22, Krasnok22, Hill23}, to the best of our knowledge they have never been observed for limit cycles as predicted in  Fig.~\ref{fig:coupled-spiking}(c).





\section{\label{sec:Conclusions}Conclusions and perspectives}
Cavities with memory in their nonlinear response, due for example to thermo-optical nonlinearities, support self-oscillations under constant driving. These limit cycles play a crucial role in excitability, which is fundamental to spiking systems like neurons. We have observed that when memory and dissipation times are commensurate, a single-mode resonator can function as an artificial neuron (AN). In this regime, the resonator exhibits Hopf and homoclinic bifurcations, corresponding to class II excitability and class I spiking, respectively. However, we have shown that considering longer or shorter memory times pose major challenges to experimental realizations.

To expand the range of memory times over which spiking behavior can be realized, we propose to couple two of the aforementioned resonators. This coupled-cavity system support limit cycles that spontaneously break mirror symmetry, and enable spike trains akin to those observed in neurons to emerge. While limit cycles can also occur in instantaneously coupled systems, they necessitate greater coupling strength and input power. Thanks to the non-instantaneous nonlinearity, spiking states persist even in the regime of weak coupling. Furthermore, our findings indicate that the threshold for driving amplitude required to trigger limit cycles increases with coupling strength, allowing us to observe spiking at lower driving amplitudes. Staying in the weak coupling regime thus makes our simple coupled cavity system more energy-efficient compared to instantaneous counterparts, which is important for neuromorphic computing applications.

Given that limit cycles manifest in states of broken symmetry, we have shown that we can excite the system either by modulating the power imbalance or by introducing random fluctuations in the laser intensity. In the latter scenario, these fluctuations induce spontaneous transitions between the two asymmetric spiking states, resulting in spontaneous symmetry breaking within the limit cycles. In other words, the system spontaneously toggles between two asymmetric states, each associated with oscillating light intensities within the resonators.

Remarkably, limit cycles in our system can occur for arbitrarily long memory times without necessitating significant adjustments to detuning or driving amplitude. One could generate long-period spike trains by employing materials characterized by thermo-optical nonlinearity and extended thermal relaxation times, as proposed in Refs.~\onlinecite{geng2020, peters2021}. This holds the promise of achieving a substantial separation of timescales within the system, with potential applications in critical systems~\cite{buendia_self-organized_2020, plenz_self-organized_2021}. The proposed setup opens the door to the development of energy-efficient artificial neural networks for all-optical information processing and high-bandwidth communication. As future research direction, we think that assessing the cascadability~\cite{vaerenbergh_cascadable_2012, wen_all-optical_2023} of our artificial neurons is important, as therein likes the key to the rational design of larger artificial neural networks.

\begin{acknowledgments}
This work was supported by the Fonds pour la Formation à la Recherche dans l’Industrie et dans l’Agriculture (FRIA) and by the Fonds National de Recherche Scientifique (FNRS) in Belgium. Computational resources have been provided by the Consortium des Équipements de Calcul Intensif (CÉCI), funded by the FNRS under Grant No.\ 2.5020.11 and by the Walloon Region. This work is part of the research programme of the Netherlands Organisation for Scientific Research (NWO).  S.R.K.R. acknowledges an ERC Starting Grant with project number 852694.
\end{acknowledgments}

\appendix



\section{\label{app:symmetric-solution} Symmetric fixed points}

Here we show that the symmetrically driven cavities support symmetric fixed points similar to the one of the single cavity. Neglecting noise, fixed points of the single cavity satisfy
\begin{equation}
    0 = \left[i(\Delta  - U |\alpha|^2) - \frac{\Gamma}{2} \right] \alpha + \sqrt{\kappa_\mathrm{L}}F
\end{equation}
which is clearly similar to Eq.~\eqref{eq:sym-solutions} giving symmetric fixed points for the coupled system. The only difference is that, for coupled cavities, the detuning $\Delta$ is shifted by the coupling $J$.

By taking the modulus squared of both side of Eq.~\eqref{eq:sym-solutions}, one sees that fixed points are roots of a third-order polynomial in $n=|\alpha|^2$,
\begin{equation}
\begin{split}
    \mathcal{P}(n) \equiv &U^2 n^3 - 2 (\Delta+J) U n^2\\
    &- \left[ (\Delta+J)^2 \frac{\Gamma^2}{4} + \right] n - \kappa_\mathrm{L} |F|^2.
\end{split}
\end{equation}
As $n \in \mathbb{R}$, $\mathcal{P}$ has either one or three roots (in between saddle-node bifurcations).

\section{\label{app:asym-conditions} Range of parameters for asymmetric solutions}

Here we give the constraints for observing asymmetric solutions under symmetric driving. All fixed points of the coupled system satisfy
\begin{equation}
\begin{split}
        &\left[i(\Delta -J  - U |\alpha_1|^2) - \frac{\Gamma}{2} \right] \alpha_1\\
        &= \left[i(\Delta - J  - U |\alpha_2|^2) - \frac{\Gamma}{2} \right] \alpha_2.
\end{split}
\end{equation}
Taking modulus squared of both sides leads to Eq.~\eqref{eq:asym-solutions}. Here the equation involves the frequency $\Delta-J$ because we are interested in asymmetric modes. Asymmetric fixed points can not cancel the first parenthesis of Eq.~\eqref{eq:asym-solutions} because $n_1\neq n_2$ by definition of the asymmetry. Therefore the only possibility for asymmetric fixed points is that the second parenthesis (the one multiplied by $(n_1-n_2)$ in Eq.~\eqref{eq:asym-solutions} cancels. The beforementioned condition means that
\begin{equation}
\begin{split}\label{eq:poly-asym}
        U^2(n_1^2+n_2^2+n_1 n_2) &- 2U(\Delta-J)(n_1+n_2)\\
        &+\frac{\Gamma^2}{4} + (\Delta - J)^2 = 0,
\end{split}
\end{equation}
that can be seen as a second-order polynomial in $n_2$ with roots
\begin{equation}\label{eq:n2}
\begin{split}
        n_2 = &\dfrac{-U n_1 + 2 (\Delta-J)}{2U}\\
        &\pm \dfrac{\sqrt{-3U^2 n_1^2 + 4U n_1(\Delta-J) - \Gamma^2}}{2 U}.
\end{split}
\end{equation}
As $n_2 \in \mathbb{R}^+$, the square root argument must be positive. This argument is a second-order polynomial in $n_1$ and is positive if $n_1$ lies between its roots
\begin{equation}\label{eq:n1}
\begin{split}
        n_1 &= \dfrac{2 (\Delta - J) \pm \sqrt{4(\Delta-J)^2 - 3\Gamma^2}}{3 U}.
\end{split}
\end{equation}
These values are reals and positives as long as ${\Delta \geq J+\sqrt{3}\Gamma/2}$ which is the condition Eq.~\eqref{eq:detuning_condition}. Thus $n_2$ in Eq.~\eqref{eq:n2} is real as long as $n_1$ is in between the roots Eq.~\eqref{eq:n1}. Because $n_1$ and $n_2$ can be swapped in Eq.~\eqref{eq:poly-asym}, it leads to condtions Eqs.~[\eqref{eq:n1_condition1}-\eqref{eq:n1_condition2}].

\section{\label{app:pitchfork} Pitchfork bifurcations}

Pitchfork bifurcations are located at the intersections of symmetric and asymmetric solutions. Therefore fixed points associated with these bifurcations are roots of Eq.~\eqref{eq:poly-asym} but also satisfy $n_1=n_2=n$. These two observations lead to
\begin{equation}
        3 U^2 n^2 - 4U(\Delta-J)n +\frac{\Gamma^2}{4} + (\Delta - J)^2 = 0.
\end{equation}
At the pitchfork bifurcations the intensity is
\begin{equation}
        n_\pm = \dfrac{2(\Delta-J) \pm \sqrt{(\Delta - J)^2 - 3\Gamma^2/4}}{3 U}.
\end{equation}
The input power corresponding to these intensities can be computed easily injecting $n_\pm$ in the modulus square of Eq.~\eqref{eq:sym-solutions}.



\bibliography{references}

\end{document}